\documentclass[11pt]{article}
\usepackage{hyperref}
\pdfoutput=1
\begin{document}
\title{Suspension flows past bluff bodies: \\ 
Investigation in a microfluidic environment}
\author{Shahab Shojaei-Zadeh$^1$ and Jeffrey F. Morris$^2$\\
\\\vspace{6pt} $^1$Mechanical and Aerospace Engineering \\
 Rutgers University \\ Piscataway, NJ 08854-8058 USA 
$^2$
Levich Institute and Chemical Engineering \\ City College of New York\\ New York, NY 10031 USA}
\maketitle
\begin{abstract}
This sequence of fluid dynamics videos illustrates the behavior of a suspension of noncolloidal particles flowing past 
various bluff body obstacles within a microfluidic device.    The polystyrene particles, of 7 $\mu$m diameter and volume fraction of 8.4\%, are carefully made neutrally buoyant with the suspending liquid composed of a mixture of water and a small fraction of glycerol.   The channel depth is 60 $\mu$m and the typical length of the obstacles normal to the flow direction is 200 $\mu$m.  The flow rate is varied to generate Reynolds numbers based on the scale of the obstacle in the approximate range $60 < Re < 500$; the narrow dimension 
in the depth direction suppresses onset of unsteadiness and vortex shedding, so that the flows studied are found to be steady (aside from particle-scale fluctuations).  Particles are observed to be depleted in the wake region of the obstacle.  In certain cases, the entire wake is clear of particles; in other cases there is a portion of the 
wake in which particles recirculate while a portion of the wake is completely devoid of particles.   Experimental observations reveal that if particles are forced into an
initially particle-depleted region, they will eventually leave and will bring the wake to its original state, implying these are steady-state distributions.  
\end{abstract}
The behavior of particle-laden flow under conditions of significant inertia is poorly understood.   Here, the classic 
flow problem of flow past a bluff body obstacle is studied for a suspension of neutrally-buoyant particles to assess the
distribution of particles in the wake region.  We use a microfluidic apparatus for this purpose, as it allows ready 
implementation of the experimental concept with ease of visualization by high-speed video imaging.    

Microfluidic channels of depth 60 $\mu$m, width 465 $\mu$m, and length of several mm were manufactured. In the middle of each channel, a cylindrical obstacle extending across the entire depth was made as part of the manufacturing process.   The generator of the cylinder was varied over a range of shapes, including in the cases illustrated here a square (of side-length $D= 200$ $\mu$m), blade with flat side to flow (200 $\mu$m by 40 $\mu$m), triangle with flat to flow, reverse triangle with point into flow.  In all cases, the cross-stream dimension is $D = 200$ $\mu$m, about 30 times the diameter of the 7 $\mu$m particles.  The particles are suspended in a mixture of water and glycerine matched to the density of 1.05 g cm$^{-3}$ of the polystyrene particles, with the liquid viscosity quite low, about $\eta = 1.8$ cP.     

The flow was imposed at a range of rates yielding Reynolds number of $Re = \rho U D/\eta$ with $U$ the mean axial speed in the range $60<Re <500$. The flow is visualized under an upright microscope, typically at 10$^4$ frames/second. The flow is found to be steady on large scales, owing to the suppression of vortex shedding by the shallow depth.   The flow determined by tracking of particles is very similar to the pure fluid flow at equivalent Reynolds numbers (accounting for added viscous dissipation caused by the suspended particles), as determined by numerical solution of the pure fluid flow in the 
geometry of the experiment using COMSOL.  Hence, the particle fraction is sufficiently low that non-Newtonian effects are largely negligible.  

Considering \href{URL of video}{the Video}, the key observation related to the two-phase aspect of the flow is that, quite generally, there is a portion of the wake which is completely devoid of particles.  In certain cases, 
namely the flow past the square, circular object, or streamline objects, the wake was essentially completely devoid of particles. 
For the blade geometry, and the triangular shapes in either orientation, a part of the recirculating wake contains particles; this region adjoins the separatrix of the wake region and the flow bypassing the wake.  Particles are believed to enter at the leading edge of the obstacle, recirculate some number of times and exit, developing a steady state through a balance of entering and exiting particles.

The steadiness of the distribution is confirmed by disturbing the wake with an injection of particles and observing the subsequent dynamics.  We have injected into the channel a bolus of material at much larger solid fraction. Upon passage of this bolus of material by the obstacle, it causes an opening of the closed streamline wake, and many particles enter.  These particles were then tracked and found to circulate a varying number of times in the wake, exiting always at the hyperbolic point where the 
bypassing flow joins behind the wake.  Note that a `weld line' depleted in solids is seen here, probably due to the basic displacement of particles relative to liquid upon interaction with the leading edge of the obstacle. 
We emphasize that the particles are carefully density matched so that the behavior seen is not due to 
differential density effects which might lead to centrifugal separation, for example.

Further details of the study, including a qualitative discussion of the mechanism believed to be responsible for 
the depletion in the wake, see the reference Shojaei-Zadeh \& Morris (2010) available from the authors. 
\par\vspace{0.2in}

\hrule\par\vspace{0.2in}
{\bf Reference:}\par
\medskip 

Shojaei-Zadeh, S. \& Morris, J. F. 2010 Probing particle transport in closed streamline flows with microfluidic devices. Submitted to {\em Phys. Fluids}.

%
\end{document}